\documentclass[aps,pre,nofootinbib,showpacs,tightenlines,preprint,titlepage,amsmath]{revtex4}
\usepackage{bm}

\newcommand{\be}{\begin{equation}}
\newcommand{\ee}{\end{equation}}
\newcommand{\bea}{\begin{eqnarray}}
\newcommand{\eea}{\end{eqnarray}}
\newcommand{\cC}{\ensuremath{\mathcal{C}}}
\newcommand{\cP}{\ensuremath{\mathcal{P}}}
\newcommand{\cT}{\ensuremath{\mathcal{T}}}

\newcommand{\half}{\mbox{$\textstyle{\frac{1}{2}}$}}
\newcommand{\e}{{\rm e}}

\begin{document}
\title{$\cP\cT$-Symmetric Quantum Electrodynamics}

\author{Carl M. Bender}\thanks{Permanent address: Department of Physics,
Washington University, St. Louis, MO 63130, USA}
\email{cmb@wustl.edu}
\affiliation{Theoretical Physics, Blackett Laboratory, Imperial College, London
SW7 2BZ, UK}

\author{Ines Cavero-Pelaez}
\email{cavero@nhn.ou.edu}
\author{Kimball A. Milton}
\email{milton@nhn.ou.edu}
\author{K. V. Shajesh}
\email{shajesh@nhn.ou.edu}

\affiliation{Oklahoma Center for High Energy Physics and Department of
Physics and Astronomy, University of Oklahoma, Norman, OK 73019, USA}

\date{\today}

\begin{abstract} 
The Hamiltonian for quantum electrodynamics becomes non-Hermitian if the
unrenormalized electric charge $e$ is taken to be imaginary. However, if one
also specifies that the potential $A^\mu$ in such a theory transforms as a
pseudovector rather than a vector, then the Hamiltonian becomes $\cP\cT$
symmetric. The resulting non-Hermitian theory of electrodynamics is the analog
of a spinless quantum field theory in which a pseudoscalar field $\varphi$ has a
cubic self-interaction of the form $i\varphi^3$. The Hamiltonian for this cubic
scalar field theory has a positive spectrum, and it has recently been
demonstrated that the time evolution of this theory is unitary. The proof of
unitarity requires the construction of a new operator called $\cC$, which is
then used to define an inner product with respect to which the Hamiltonian is
self-adjoint. In this paper the corresponding $\cC$ operator for non-Hermitian
quantum electrodynamics is constructed perturbatively. This construction
demonstrates the unitarity of the theory. Non-Hermitian quantum electrodynamics
is a particularly interesting quantum field theory model because it is
asymptotically free.
\end{abstract}

\maketitle
\section{Introduction}
\label{sec1}
It is common wisdom that the Hamiltonian that defines a quantum theory should be
Hermitian $H=H^\dag$, where the symbol $\dag$, which indicates Dirac Hermitian
conjugation, represents the combined operations of complex conjugation and
matrix transposition. There are two reasons given for requiring that the
Hamiltonian be Hermitian: First, the condition $H=H^\dag$ guarantees that the
energy eigenvalues of $H$ will be real. Second, this condition guarantees that
time evolution will be unitary; that is, that probability will be conserved.

However, in the past few years it has become clear that the requirements of
spectral positivity and unitarity can be met even if the Hamiltonian is not
Hermitian in the Dirac sense. The first non-Hermitian Hamiltonian for which
these two properties were verified was the quantum-mechanical model
\be
H=p^2+x^2(ix)^\epsilon\quad(\epsilon\geq0).
\label{e1}
\ee
It was observed in 1998 that the spectrum of this class of Hamiltonians was
positive and discrete \cite{BB} and it was conjectured that spectral positivity
was a consequence of the invariance of $H$ under the combination of the
space-reflection operator $\cP$ and the time-inversion operator $\cT$. Three
years later, a proof of spectral positivity was given \cite{DDT}. Then, in 2002
it was shown that the Hamiltonian in (\ref{e1}) defines a unitary time
evolution \cite{BBJ}. Specifically, it was demonstrated that if the $\cP\cT$
symmetry of a non-Hermitian Hamiltonian is unbroken, then it is possible to
construct a new operator called $\cC$ that commutes with the Hamiltonian $H$.
The Hilbert space inner product with respect to the $\cC\cP\cT$ adjoint has a
positive norm and the time evolution operator $\e^{iHt}$ is unitary. Thus, from
this quantum-mechanical study it is clear that Dirac Hermiticity of the
Hamiltonian is not a necessary requirement of a quantum theory; unbroken $\cP
\cT$ symmetry is sufficient to guarantee that the spectrum of $H$ is real and
positive and that the time evolution is unitary.
 
The construction of the $\cC$ operator in Ref.~\cite{BBJ} was the key step in
showing that the non-Hermitian Hamiltonian (\ref{e1}) exhibits unitary time
evolution. However, the difficulty with the construction given in
Ref.~\cite{BBJ} is that the calculation of the $\mathcal{C}$ operator required
as input all the coordinate-space eigenvectors of the Hamiltonian. While this
information is, in principle, available in quantum mechanics, it is hardly
available for a quantum field theory because there is no simple analog of the
coordinate-space Schr\"odinger equation. Thus, the analysis in Ref.~\cite{BBJ}
does not extend easily to quantum field theory.

However, it was recently shown that a perturbative construction of $\cC$ that
does not require the eigenfunctions of the Hamiltonian is possible for the case
of a scalar quantum field theory with a cubic self-interaction of the form $i
\phi^3$ \cite{BBJ04}. This result is particularly important because this quantum
field theory has already appeared in the literature in studies of the Lee-Yang
edge singularity \cite{r11} and in Reggeon field theory \cite{r10}. The
construction of the $\cC$ operator for the $i\phi^3$ field theory shows that
this quantum field theory is a fully acceptable unitary quantum theory and not
just an interesting but unrealistic mathematical curiosity.

Furthermore, an exact construction of the $\cC$ operator \cite{LEE} was carried
out for the Lee model, a cubic quantum field theory in which mass,
wave-function, and coupling-constant renormalization can be done exactly
\cite{lee}. The construction of the $\cC$ operator for the Lee model explains a
long-standing puzzle. It is known that there is a critical value of the
renormalized coupling constant $g$ for the Lee model and that when $g$ exceeds
this critical value, the unrenormalized coupling constant becomes pure
imaginary, and hence the Hamiltonian becomes non-Hermitian. As a consequence,
a ghost state having negative Hermitian norm appears when $g>g_{\rm crit}$, and
the presence of this ghost state causes the $S$ matrix to be nonunitary. By
constructing the $\cC$ operator we can reinterpret the Hilbert space for the
theory. By using a $\cC\cP\cT$ inner product, the ghost state now has a positive
norm and the Lee model becomes a consistent unitary quantum field theory. This
physical reinterpretation of the Lee model was anticipated by F.~Kleefeld in a
beautiful series of papers \cite{K}.

Recently, additional progress was made in understanding the $\cC$ operator in
the context of an $ig\phi^3$ quantum field theory. It was shown that $\cC$
transforms as a scalar under the action of the homogeneous Lorentz group
\cite{scalar}. In this paper it was argued that because the Hamiltonian $H_0$
for the unperturbed theory ($g=0$) commutes with the parity operator $\cP$, the
intrinsic parity operator $\cP_{\rm I}$ in the noninteracting theory transforms
as a Lorentz scalar. (The {\it intrinsic} parity operator $\cP_{\rm I}$ and the
parity operator $\cP$ have the same effect on the fields, except that $\cP_{\rm
I}$ does not reverse the sign of the spatial argument of the field.) When the
coupling constant $g$ is nonzero, the parity symmetry of $H$ is broken and $\cP_
{\rm I}$ is no longer a scalar. However, $\cC$ {\it is} a scalar. Since $\lim_{g
\to0}\cC=\cP_{\rm I}$, one can interpret the $\cC$ operator as the complex
extension of the intrinsic parity operator when the imaginary coupling constant
is turned on.

In this paper we examine $\cP\cT$-symmetric quantum electrodynamics, a
non-Hermitian quantum field theory that is much more interesting than an $i\phi^
3$ field theory. Unlike the scalar $i\phi^3$ field theory, $\cP\cT$-symmetric
quantum electrodynamics possesses many of the features of conventional quantum
electrodynamics, including Abelian gauge invariance. Two earlier preliminary
studies of this theory have already been published \cite{pre1,pre2}. The advance
reported in the present paper is the construction of the $\cC$ operator to
leading order in perturbation theory for this remarkable theory. Our
construction provides strong evidence that $\cP\cT$-symmetric quantum
electrodynamics is a viable and consistent unitary quantum field theory.

While $\cP\cT$-symmetric quantum electrodynamics is similar to an $i\phi^3$
field theory because its interaction Hamiltonian is cubic and its coupling
constant is pure imaginary, this quantum field theory is especially interesting
because, like a $\cP\cT$-symmetric $-\phi^4$ scalar quantum field theory in four
dimensions, $\cP\cT$-symmetric electrodynamics is asymptotically free
\cite{asym}. The only asymptotically free quantum field theories described by
Hermitian Hamiltonians are those that possess a {\it non-Abelian} gauge
invariance; $\mathcal{PT}$ symmetry allows for new kinds of asymptotically free
theories that do not have to possess a non-Abelian gauge invariance.

\section{$\mathcal{PT}$-symmetric quantum electrodynamics}
\label{sec2}

In order to formulate a Lorentz covariant quantum field theory one begins by
specifying the Lorentz transformation properties of the quantum fields under the
proper orthochronous Lorentz group. [For example, one can specify that the field
$\phi({\bf x},t)$ transforms as a scalar.] In addition, one is free to specify
the transformation properties of the fields under parity reflection. [For
example, one can specify that $\phi({\bf x},t)$ transforms as a scalar, so that
it does not change sign under $\cP$, or that it transforms as a pseudo-scalar,
so that it changes sign under $\cP$.] Having fully specified the transformation
properties of the fields, one then formulates the (scalar) Lagrangian in terms
of these fields.

A non-Hermitian but $\mathcal{PT}$-symmetric version of electrodynamics can be
constructed by assuming that the four-vector potential transforms as an {\it
axial} vector \cite{pre2}. As a consequence, the electromagnetic fields
transform under parity reflection like
\be
\mathcal{P}:\quad {\bf E\to E},\quad {\bf B\to-B}, \quad {\bf A\to A},
\quad A^0\to-A^0.
\label{e2}
\ee
Under time reversal, the transformations are assumed to be conventional:
\be
\mathcal{T}:\quad {\bf E\to E},\quad {\bf B\to- B}, \quad {\bf A\to -A},
\quad A^0\to A^0.
\label{e3}
\ee
The Lagrangian of the theory then possesses an imaginary coupling constant in
order that it be invariant under the product of these two symmetries:
\be
\mathcal{L}=-\textstyle{\frac{1}{4}}F^{\mu\nu}F_{\mu\nu}+\half\psi^\dagger\gamma
^0\gamma^\mu\frac1i\partial_\mu\psi+\half m\psi^\dagger\gamma^0\psi+ie\psi^\dag
\gamma^0\gamma^\mu\psi A_\mu.
\label{e4}
\ee
The corresponding Hamiltonian density is then
\be
\mathcal{H}=\half(E^2+B^2)+\psi^\dagger\left[\gamma^0\gamma^k\left(-i\nabla_k+ie
A_k\right)+m\gamma^0\right]\psi.
\label{e5}
\ee
The electric current appearing in both the Lagrangian and Hamiltonian densities,
$j^\mu=\psi^\dagger\gamma^0\gamma^\mu\psi$, transforms conventionally under both
$\mathcal{P}$ and $\mathcal{T}$:
\begin{subequations}
\bea
\mathcal{P}j^\mu({\bf x},t)\mathcal{P}&=&\left(\begin{array}{c}
j^0\\-{\bf j}\end{array}\right)(-{\bf x},t),\label{5a}\\
\mathcal{T}j^\mu({\bf x},t)\mathcal{T}&=&\left(\begin{array}{c}
j^0\\-{\bf j}\end{array}\right)({\bf x},-t).
\label{5b}
\eea
\end{subequations}

Just as in the case of ordinary quantum electrodynamics, $\cP\cT$-symmetric
electrodynamics has an Abelian gauge invariance. In this paper we choose to work
in the Coulomb gauge, $\bm{\nabla}\cdot{\bf A}=0$, so the nonzero canonical
equal-time commutation relations are
\begin{subequations}
\bea
\{\psi_a({\bf x},t),\psi_b^\dagger({\bf y},t)\}&=&\delta_{ab}\delta({\bf x-y}),
\label{6a}\\
{}[A_i^T(\mathbf{x}),E_j^T(\mathbf{y})]&=&-i\left[\delta_{ij}-\frac{\nabla_i
\nabla_j}{\nabla^2}\right]\delta(\mathbf{x-y}),
\label{6b}
\eea
\end{subequations}
where $T$ denotes the transverse part,
\be
\bm{\nabla}\cdot \mathbf{A}^T=\bm{\nabla}\cdot \mathbf{E}^T=0.
\ee
In the following, the symbols $\mathbf{E}$ and $\mathbf{B}$ represent the
transverse parts of the electromagnetic fields, so
\be
\bm{\nabla}\cdot \mathbf{E}=\bm{\nabla}\cdot \mathbf{B}=0.
\label{trans}
\ee

\section{Calculation of the $\cC$ Operator}
\label{sec3}

As in quantum-mechanical systems and scalar quantum field theories, we seek a
$\cC$ operator in the form \cite{BBJ04}
\be
\cC=\e^Q\cP,
\label{7}
\ee
where $\cP$ is the parity operator, and our objective will be to calculate
the operator $Q$ \cite{tech}. Because $\mathcal{C}$ must satisfy the three
defining properties
\begin{subequations}
\bea
\mathcal{C}^2&=&1,
\label{8a}\\
{}[\mathcal{C},\mathcal{PT}]&=&0,
\label{8b}\\
{}[\mathcal{C},H]&=&0,
\label{8c}
\eea
\end{subequations}
we infer from Eq.~(\ref{8a}) that 
\begin{subequations}
\be
Q=-\mathcal{P}Q\mathcal{P},
\label{9a}
\ee
and because $\mathcal{PT}=\mathcal{TP}$, we infer from (\ref{8b}) that
\be
Q=-\mathcal{T}Q\mathcal{T}.
\label{9b}
\ee
\end{subequations}
The two equations (\ref{8a}) and (\ref{8b}) can be thought of as kinematical
constraints on $Q$.

The third equation (\ref{8c}), which can be thought of as a dynamical
condition on $Q$, allows us to determine $Q$ perturbatively.
If we separate the interaction part of the Hamiltonian from the free part,
\be
H=H_0+eH_1,
\label{10}
\ee
and seek $Q$ in the form of a power series
\be Q=eQ_1+e^2 Q_2+\cdots,
\label{11}
\ee
then the first contribution to the $Q$ operator is determined by
\be
[Q_1,H_0]=2H_1.
\label{12}
\ee
As in previous studies of cubic quantum theories, the second correction commutes
with the Hamiltonian,
\be
[Q_2,H_0]=0,
\label{13}
\ee
and Eq.~(\ref{11}) reduces to a series in odd powers of $e$,
\be
Q=eQ_1+e^3 Q_3+\cdots,
\label{14}
\ee
which illustrates the virtue of the exponential representation (\ref{7}).

To use Eq.~(\ref{12}) to determine the operator $Q_1$, we construct the most
general nonlocal {\it ansatz} for the operator $Q_1$ in terms of the sixteen
independent Dirac tensors. There is no condition of gauge invariance on this
operator because we have chosen to work in the Coulomb gauge. There are sixteen
tensor functions in principle, which we take to be defined by 
\bea
Q_1&=&\int d\mathbf{x}\,d\mathbf{y}\,d\mathbf{z}\Bigg\{\left[
f_+^{kl}(\mathbf{x},
\mathbf{y},\mathbf{z})E^k(\mathbf{x})+f_-^{kl}(\mathbf{x},
\mathbf{y},\mathbf{z})B^k(\mathbf{x})\right]\psi^\dagger(\mathbf{y})\gamma^0
\gamma^l\psi(\mathbf{z})\nonumber\\
&&\quad\mbox{}+\left[g_+^{k}(\mathbf{x},
\mathbf{y},\mathbf{z})E^k(\mathbf{x})+g_-^{k}(\mathbf{x},
\mathbf{y},\mathbf{z})B^k(\mathbf{x})\right]\psi^\dagger(\mathbf{y})\gamma^0
\gamma^5\psi(\mathbf{z})\nonumber\\
&&\quad\mbox{}+\left[h_+^{k}(\mathbf{x},
\mathbf{y},\mathbf{z})E^k(\mathbf{x})+h_-^{k}(\mathbf{x},
\mathbf{y},\mathbf{z})B^k(\mathbf{x})\right]\psi^\dagger(\mathbf{y})\gamma^5
\psi(\mathbf{z})\nonumber\\
&&\quad\mbox{}+\left[j_+^{kl}(\mathbf{x},
\mathbf{y},\mathbf{z})E^k(\mathbf{x})+j_-^{kl}(\mathbf{x},
\mathbf{y},\mathbf{z})B^k(\mathbf{x})\right]\psi^\dagger(\mathbf{y})
\gamma^l\psi(\mathbf{z})\nonumber\\
&&\quad\mbox{}+\left[F_+^{kl}(\mathbf{x},
\mathbf{y},\mathbf{z})B^k(\mathbf{x})+F_-^{kl}(\mathbf{x},
\mathbf{y},\mathbf{z})E^k(\mathbf{x})\right]
\psi^\dagger(\mathbf{y})\gamma^0\gamma^5\gamma^l\psi(\mathbf{z})\nonumber\\
&&\quad\mbox{}+\left[G_+^{k}(\mathbf{x},
\mathbf{y},\mathbf{z})B^k(\mathbf{x})+G_-^{k}(\mathbf{x},
\mathbf{y},\mathbf{z})E^k(\mathbf{x})\right]\psi^\dagger(\mathbf{y})\gamma^0
\psi(\mathbf{z})\nonumber\\
&&\quad\mbox{}+\left[H_+^{k}(\mathbf{x},
\mathbf{y},\mathbf{z})B^k(\mathbf{x})+H_-^{k}(\mathbf{x},
\mathbf{y},\mathbf{z})E^k(\mathbf{x})\right]\psi^\dagger(\mathbf{y})
\psi(\mathbf{z})\nonumber\\
&&\quad\mbox{}+[J_+^{kl}(\mathbf{x},
\mathbf{y},\mathbf{z})B^k(\mathbf{x})+J_-^{kl}(\mathbf{x},
\mathbf{y},\mathbf{z})E^k(\mathbf{x})]\psi^\dagger(\mathbf{y})\gamma^5
\gamma^l\psi(\mathbf{z})\Bigg\}.\label{15}
\eea
In Eq.~(\ref{15}) we have taken into account the fact that the electric and magnetic
fields are transverse, $\bm{\nabla}\cdot\mathbf{E}=\bm{\nabla}\cdot\mathbf{B}=0$
[see Eq.~(\ref{trans})]. The parity constraint (\ref{9a}) is satisfied because
$f_\pm$, $g_\pm$, $\cdots$, are respectively even and odd functions:
\be
f_\pm(\mathbf{x,y,z})=\pm f_\pm (\mathbf{-x,-y,-z}).
\label{16}
\ee
We will see that the time-reversal constraint (\ref{9b}) is automatically
satisfied by $Q_1$ in (\ref{15}).

The solution of Eq.~(\ref{12}) is obtained by using the canonical
commutation relations (\ref{6a}) and (\ref{6b}), which imply that
\begin{subequations}
\bea
\left[E^k(\mathbf{x}),\frac12\int d\mathbf{w} B^2(\mathbf{w})\right]
&=&i(\bm{\nabla}\times \mathbf{B})_k(\mathbf{x}),\label{17a}\\
{}\left[B^k(\mathbf{x}),\frac12\int d\mathbf{w} E^2(\mathbf{w})\right]
&=&-i(\bm{\nabla}\times \mathbf{E})_k(\mathbf{x}),\label{17b}
\eea
\bea
&&\left[\int d\mathbf{y}\,d\mathbf{z}\,\phi(\mathbf{y,z})\psi^\dagger(
\mathbf{y})\Gamma\psi(\mathbf{z}),\int d\mathbf{w}\,\psi^\dagger(\mathbf{w})
\gamma^0\gamma^k\frac1i \nabla_k\psi(\mathbf{w})\right]\nonumber\\
&&\qquad\qquad
=\frac{i}2\int d\mathbf{y}\,d\mathbf{z}\big[(\nabla^z_k+\nabla^y_k)\phi(
\mathbf{y,z})\psi^\dagger(\mathbf{y})
\{\Gamma,\gamma^0\gamma^k\}\psi(\mathbf{z})\nonumber\\
&&\qquad\qquad\qquad\mbox{}+(\nabla^z_k-\nabla^y_k)\phi(
\mathbf{y,z})\psi^\dagger(\mathbf{y})
[\Gamma,\gamma^0\gamma^k]\psi(\mathbf{z})\big],\label{17c}\\
&&\left[\int d\mathbf{y}\,d\mathbf{z}\,\phi(\mathbf{y,z})\psi^\dagger(
\mathbf{y})\Gamma\psi(\mathbf{z}),m\int d\mathbf{w}\,\psi^\dagger(\mathbf{w})
\gamma^0\psi(\mathbf{w})\right]\nonumber\\
&&\qquad\qquad=m\int d\mathbf{y}\,d\mathbf{z}\,\phi(
\mathbf{y,z})\psi^\dagger(\mathbf{y})[\Gamma,\gamma^0]\psi(\mathbf{z}).\label{17d}
\eea
\end{subequations}
There are sixteen resulting equations for the tensor coefficients, which break
up into two independent sets of eight equations each. Since there is only one
inhomogeneous term, this means that the coefficients that satisfy the set of
equations with no driving term must vanish. The remaining equations are most
conveniently written in momentum space, where the Fourier transform is defined
by
\be
\tilde f(\mathbf{p})=\int d\mathbf{x}\,\e^{-i\mathbf{p\cdot x}}f(\mathbf{x}).
\label{18}
\ee

If the momenta corresponding to the coordinates $\mathbf{x,y,z}$ are $\mathbf{p,
q,r}$, then as a result of translational invariance there is an overall
momentum-conserving delta function, which sets $\mathbf{p+q+r=0}$. Using dyadic
notation, it is not hard to show that these equations are, in terms of the two
independent vectors $\mathbf{p}$ and $\mathbf{t=r-q}$, given by
\begin{subequations}
\bea
\mathbf{p\times \tilde g_-}+\mathbf{\tilde J_-\cdot t}
-2m\mathbf{\tilde h_+}&=&\mathbf{0},\label{19a}\\
\mathbf{p\times \tilde h_+}+\mathbf{\tilde F_+\cdot p}
+2m\mathbf{\tilde g_-}&=&\mathbf{0},\label{19b}\\
\mathbf{p\times \tilde j_-}-i\mathbf{\tilde J_-\times p}-\mathbf{\tilde G_- t}
-2m\mathbf{\tilde f_+}&=&\mathbf{0},
\label{19c}\\
\mathbf{p\times \tilde F_+}-\mathbf{\tilde h_+p}+i\mathbf{\tilde f_+\times t}
&=&\mathbf{0},\label{19d}\\
\mathbf{p\times \tilde G_-}+\mathbf{\tilde j_-\cdot t}&=&\mathbf{0},
\label{19e}\\
\mathbf{p\times \tilde J_-}-\mathbf{\tilde g_- t}+i\mathbf{\tilde j_-\times p}
&=&\mathbf{0},\label{19f}\\
\mathbf{p\times \tilde{H}_+}+\mathbf{\tilde f_+\cdot p}&=&\mathbf{0},
\label{19g}\\
\mathbf{p\times \tilde f_+}-i\mathbf{\tilde F_+\times t}-\mathbf{\tilde H_+ p}
+2m\mathbf{\tilde j_-}
&=&\frac2{p^2}\mathbf{1\times p}.\label{19h}
\eea
\end{subequations}

We may take all the coefficient tensors to be transverse to $\mathbf{p}$ in the
first index,
\be
\mathbf{p\cdot \tilde f_+}=\mathbf{0}, \quad \mathbf{p\cdot \tilde F_+}=
\mathbf{0}, \quad \mathbf{p\cdot \tilde g_-}
=0,\label{20}
\ee
and so on, which is consistent with the transversality of the electric and
magnetic fields appearing in the construction (\ref{15}) of $Q_1$.
This property then allows us to solve Eqs.~(\ref{19d}), (\ref{19e}),
(\ref{19f}) and (\ref{19g}) for $\mathbf{\tilde F_+}$,
$\mathbf{\tilde G_-}$,  $\mathbf{\tilde H_+}$, and $\mathbf{\tilde J_-}$ 
in terms 
of $\mathbf{\tilde f_+}$, $\mathbf{\tilde g_-}$, $\mathbf{\tilde h_+}$, 
and $\mathbf{\tilde j_-}$:
\begin{subequations}
\bea
\mathbf{\tilde F}_+&=&\frac1{p^2}\left(-\mathbf{p\times \tilde h_+ p}
+i\mathbf{p\times\tilde f_+\times t}\right),\label{4p}\\
\mathbf{\tilde G}_-&=&\frac1{p^2}\mathbf{p\times \tilde j_-\cdot t},\label{5p}\\
\mathbf{\tilde J}_-&=&-\frac1{p^2}\left(\mathbf{p\times \tilde g_- t}
-i\mathbf{p\times\tilde j_-\times p}\right),\label{6p}\\
\mathbf{\tilde H}_+&=&\frac1{p^2}\mathbf{p\times \tilde f_+\cdot p}.\label{7p}
\eea
\end{subequations}
The remaining four equations then imply that
\begin{subequations}
\bea
\mathbf{p\times\tilde g_-}\left(p^2-t^2\right)+i
\mathbf{p\times \tilde j_-\cdot(p\times t)}-2mp^2\mathbf{\tilde h_+}
&=&\mathbf{0},\label{1p}\\
i\mathbf{p\times \tilde f_+\cdot(p\times t)}-2mp^2\mathbf{\tilde g_-}&=&
\mathbf{0},\label{2p}\\
\mathbf{p\times \tilde j_-}\cdot\left(\mathbf{pp}-\mathbf{tt}
\right)-i\mathbf{p\times \tilde g_- p\times t}-2mp^2\mathbf{\tilde f_+}&=&
\mathbf{0},\label{3p}\\
\mathbf{p\times \tilde f_+}\cdot\left[(\mathbf{tt-1}t^2)-(\mathbf{pp-1}p^2)
\right]+i\mathbf{p\times \tilde h_+ p\times t}+2mp^2\mathbf{\tilde j_-}&=&
2(\mathbf{1\times p}).\label{8p}
\eea
\end{subequations}
Equations (\ref{2p}) and (\ref{1p}) allow us to solve immediately for
$\mathbf{\tilde g_-}$ and $\mathbf{\tilde f_+}$ in terms of 
$\mathbf{\tilde j_-}$ and $\mathbf{\tilde h_+}$:
\begin{subequations}
\bea
\mathbf{\tilde g_-}&=&\frac1{2mp^2}i\mathbf{p\times \tilde f_+\cdot(p\times t)},
\label{2pp}\\
\mathbf{\tilde h_+}&=&\frac{i}{2mp^2}\left[\mathbf{p\times \tilde j_
-\cdot(p\times t)}
+(t^2-p^2)\frac1{2m}\mathbf{\tilde f_+\cdot(p\times t)}\right],\label{1pp}
\eea
\end{subequations}
and then from Eqs.~(\ref{3p}) and (\ref{8p}) we obtain two equations for
$\mathbf{\tilde j_-}$ and $\mathbf{\tilde f_+}$:
\begin{subequations}
\bea
\mathbf{p\times \tilde j_-\cdot(tt-pp)}+2mp^2\mathbf{\tilde f_+}
\cdot\left[\mathbf{1}+
\frac{\mathbf{(p\times t)(p\times t)}}{4m^2p^2}\right]&=&\mathbf{0},\label{3pp}
\\
\mathbf{p\times \tilde f_+}\cdot\left[(\mathbf{tt-1}t^2)-(\mathbf{pp-1}p^2)+
\frac{t^2-p^2}{4m^2p^2}\mathbf{(p\times t)(p\times t)}\right]\quad&&\nonumber\\
\mbox{}+2mp^2\mathbf{\tilde j_-}
\cdot\left[\mathbf{1}+\frac{\mathbf{(p\times t)(p\times t)}}
{4m^2p^2}\right]&=&2(\mathbf{1\times p}).\label{8pp}
\eea
\end{subequations}
{}From Eq.~(\ref{3pp}) we see that
\be
\mathbf{\tilde f_+\cdot(t\times p)=0}.\label{3ppp}
\ee
Then we can solve Eq.~(\ref{3pp}) for $\mathbf{f_+}$ in terms of $\mathbf{j_-}$,
which when substituted into  Eq.~(\ref{8pp}) yields an equation that can be
solved easily for $\mathbf{j_-}$.

In this way it is straightforward to solve for all the coefficient tensors. In
terms of the denominator
\be
\Delta=4m^2p^2 +k^2,\label{21}
\ee
where $\mathbf{k=p\times t}$, the nonzero tensor coefficients in $Q_1$ are
\begin{subequations}
\bea
\mathbf{\tilde F_+}&=&\frac{2i}{p^2\Delta}\mathbf{p\times k \,p},\label{22a}\\
\mathbf{\tilde f_+}&=&-\frac{2}{p^2\Delta}\mathbf{p\times k\,t},\label{22b}\\
\mathbf{\tilde j_-}
&=&\frac{4 m}{\Delta}\mathbf{1\times p},
\label{22c}\\
\mathbf{\tilde J_-}&=&-i\mathbf{j_-},\label{22d}\\
\mathbf{\tilde h_+}&=&-\frac{2i}{\Delta}\mathbf{k},\label{22e}\\
\mathbf{\tilde H_+}&=&2\frac{\mathbf{p\cdot t}}{p^2}\frac{\mathbf{k}}{\Delta},
\label{22f}\\
\mathbf{\tilde g_-}&=&\mathbf{0},\label{22g}\\
\mathbf{\tilde G_-}&=&\frac{4m}{p^2\Delta}\mathbf{p\times k}.\label{22h}
\eea
\end{subequations}
Note that the parity constraint (\ref{16}) is satisfied because the $+$
quantities are even under $\mathbf{p\to-p}$, $\mathbf{t\to-t}$, while the
$-$ quantities are odd.  The time-reversal constraint (\ref{9b}) is satisfied
because of the presence of $i$ in $\mathbf{\tilde F_+}$, $\mathbf{\tilde J_-}$,
and $\mathbf{\tilde h_+}$, owing to $\mathcal{T}$ being an antiunitary operator.
The odd functions undergo another sign change under $\mathcal{T}$ because all
momenta change sign [see Eq.~(\ref{18})]. 
\section{Conclusions}
By constructing the first-order term in the $Q$ operator and thus the leading
approximation to the $\mathcal{C}$ operator, we have provided convincing
evidence that the $\mathcal{PT}$-symmetric quantum electrodynamics originally
proposed in Ref.~\cite{pre2} is unitary and that this construction enables us to
obtain a unitary $S$ matrix for the theory. Therefore, there can be little doubt
that such a $\mathcal{PT}$-symmetric theory is self-consistent and one should
now investigate whether such a theory may be used to describe natural phenomena.
Indeed, this theory provides an interesting test of Gell-Mann's {\it
Totalitarian Principle}, which states that ``Everything which is not forbidden
is compulsory'' \cite{Gell}.

\acknowledgments{CMB is grateful to the Theoretical Physics Group at Imperial
College for its hospitality and he thanks the U.K. Engineering and Physical
Sciences Research Council and the John Simon Guggenheim Foundation for their
support. We thank the U.S. Department of Energy for partial financial support of
this work.}


\begin{thebibliography}{99}
\bibitem{tech} In Refs. \cite{LEE,scalar} it was shown that the correct
representation of the $\cC$ operator has the form $\cC=e^Q\cP_{\rm I}$, where 
$\cP_{\rm I}$ is the {\it intrinsic} parity reflection operator. (The difference
between $\cP$ and $\cP_{\rm I}$ is that $\cP_{\rm I}$ does not reflect the
spatial arguments of the fields.) However, this is a technical distinction for
the case of a cubic interaction Hamiltonian because it does not affect the final
result for the $Q$ operator.
\end{thebibliography}
\end{document}